%% file: hicss51.tex
\documentclass[10pt]{article}
\input{hicss51-packages.tex}

\usepackage{booktabs} 
\usepackage[capitalise,noabbrev]{cleveref} 
\usepackage{tabularx}
\usepackage{xcolor,colortbl}
\usepackage{comment}
\usepackage{enumitem}

\usepackage[firstpage=true, placement=bottom, color=black]{background} 
\usepackage{tikz}
\newcommand\copyrighttext{%
 \footnotesize
 This is the author's version of the work. It is posted here for your personal use. The definitive Version of Record is:
 
 \smallskip
 
 Christoph Matthies and Franziska Dobrigkeit, ``Towards Empirically Validated Remedies for Scrum  Retrospective  Headaches,''  in \\\emph{Proceedings of the 53rd Hawaii International Conference on System Sciences}, 2020. http://hdl.handle.net/10125/64504 (CC BY-NC-ND 4.0)
 }
\newcommand\copyrightnotice{%
\begin{tikzpicture}[remember picture, overlay]
\node[anchor=south, yshift=10pt] at (current page.south) {\fbox{\parbox{\dimexpr\textwidth-\fboxsep-\fboxrule\relax}{\copyrighttext}}};
\end{tikzpicture}%
}
\backgroundsetup{opacity=1, scale=1, angle=0, contents={%
\copyrightnotice
}}

\title{Towards Empirically Validated Remedies\\for Scrum Retrospective Headaches}

\author{Christoph Matthies \\
  Hasso Plattner Instititute, \\
  University of Potsdam \\
  {\underline{christoph.matthies@hpi.de}} \\\And
  Franziska Dobrigkeit \\
  Hasso Plattner Instititute,\\ 
  University of Potsdam \\
  {\underline{franziska.dobrigkeit@hpi.de} }\\
 }
\date{}

\newcommand{\boxbox}[2]{
    \begin{center}
    \fbox{
        \parbox{#1\columnwidth}{#2
        }
    }
    \end{center}
}

\newcommand{\multicell}[2][t]{\begin{tabular}[#1]{@{}l@{}}#2\end{tabular}} 

\definecolor{warning}{RGB}{255,240,165}
\definecolor{success}{RGB}{198,255,165}
\definecolor{error}{RGB}{255,171,165}
\definecolor{info}{RGB}{206,255,253}

\begin{document}
\maketitle

\begin{abstract}
\input{abstract.tex}
\end{abstract}

\input{content.tex}

\newpage



\bibliographystyle{ieeetr}


\bibliography{library_retros}

\end{document}

%% file: hicss51-packages.tex
\usepackage[letterpaper]{geometry}
\usepackage{hicss51}
\usepackage{times}
\usepackage[none]{hyphenat}
\usepackage{url}
\usepackage{latexsym}
\usepackage{indentfirst}
\usepackage{graphicx}
\graphicspath{{images/}}

%% file: abstract.tex
Agile methods, especially Scrum, have become staples of the modern software development industry. Retrospective meetings are Scrum's instrument for process improvement and adaptation. They are considered one of the most important aspects of the Scrum method and its implementation in organizations. However, Retrospectives face their own challenges. Agile practitioners have highlighted common problems, i.e. headaches, that repeatedly appear in meetings and negatively impact the quality of process improvement efforts. To remedy these headaches, Retrospective activities, which can help teams think together and break the usual routine, have been proposed. In this research, we present case studies of educational and industry teams, investigating the effects of  eleven Retrospective activities on five identified headaches. While we find evidence for the claimed benefits of activities in the majority of studied cases, application of remedies also led to new headaches arising.

%% file: content.tex
\section{Introduction}
\begin{figure}[htb]
    \centering
	\includegraphics[width=\linewidth]{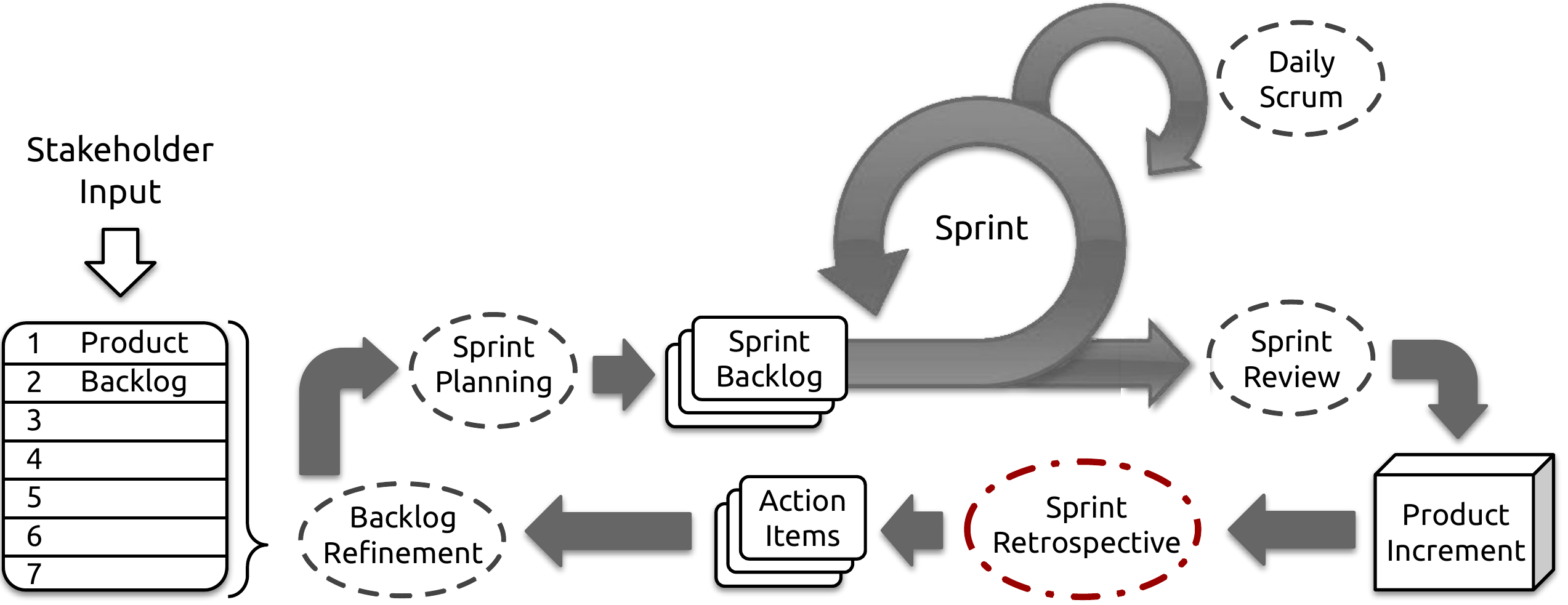}
	\caption{Scrum process flow, with the Retrospective meeting highlighted, based on~\cite{Matthies2018b}.}
	\label{fig:process}
\end{figure}

The software development industry has embraced the use of modern Agile development methods, in particular, Scrum.
The 13th Annual State of Agile Report, which analyzed 1,319 survey responses from industry practitioners, found Scrum and Scrum / XP Hybrids to be the most commonly employed Agile methods, with 64\% of respondents' organizations using them in projects~\cite{StateOfAgile2019}.
Concerning Agile technique selection, running Agile Retrospective meetings (\emph{Retros}, for short)  was found to be prominent (80\%), with only the Daily Standup, another practice focused on feedback, being more popular (86\%)~\cite{StateOfAgile2019}.
The 2017 / 2018 State of Scrum Report, a survey of more than 2,000 Agile Alliance members in 91 countries, notes that 97\% of current Scrum and Agile users indicated their commitment to continue using the methods, and 85\% reported that Scrum had improved the quality of their work life~\cite{ScrumAlliance2018}.
The report similarly highlights the importance of feedback and continuous improvement within modern development processes: 81\% of respondents held a Retro after each Sprint, while 87\% ran a Daily Scrum meeting~\cite{ScrumAlliance2018}.
The generalized flow through the Scrum method and the role of the Retros are depicted in \Cref{fig:process}.

\subsection{Agile Retrospective Meetings}
The published literature agrees with Agile practitioners' assessments regarding the importance of inspect-and-adapt cycles and reflection in Agile methods.
Retros are considered a vital component of Scrum and its implementation in organizations~\cite{dingsoyr2018}.
In fact, in the second edition of his influential book on Scrum and XP, Kniberg explicitly acknowledges this, stating that ``the retrospective is the number-one-most-important thing in Scrum''~\cite{Kniberg2015}.
The Scrum Guide defines the purpose of Retros as inspecting ``how the last Sprint went with regards to people, relationships, process, and tools''~\cite{Schwaber2017}.
Thus, Retros not only represent opportunities for teams to improve their ways of collaboration and teamwork, but also empowerment and enjoyment in future development iterations~\cite{Derby2006}.
As a part of the Scrum framework, Retros are performed at the end of an iteration, following the Sprint Review.
Outcomes of these meetings are a list of ``identified improvements''~\cite{Schwaber2017} or ``action items''~\cite{Derby2006} that will be enacted in the following Sprint.

While the Sprint Review meeting focuses on \emph{what} was produced during the last iteration, i.e. its quality and completeness, the Retro deals with \emph{how} it was built and how the next Sprint can be improved, i.e. the process and the human factors involved.

\subsection{Retrospective Headaches}
Retros have their own set of requirements for participants and face their own set of challenges.
In particular, the literature concerning Agile practitioners has highlighted problems which repeatedly appear in these meetings~\cite{rubin12,kua2013}.
These \emph{headaches} of process facilitators refer to problems of organization, process, engagement and collaboration within Retros of teams.
Recurring headaches, while not fatal to the executed development process as a whole, can negatively influence the effectiveness of Retros and the quality of Retro outcomes and process improvement efforts in general.
The identified Retro headaches include problems such as too little preparation by process facilitators~\cite{kua2013}, resulting in frustrated participants or participants not voicing their opinions, which diminishes the output of Retros~\cite{rubin12}.

\subsection{Retrospective Remedies}
To alleviate Retro headaches, multiple types of remedies have been proposed by both researchers and practitioners~\cite{Derby2006,kua2013,goncalves2014,caroli2015}.
These remedies often take the form of games or timeboxed group activities that suggest specific meeting actions for Retros and can help teams ``think together''~\cite{Derby2006} and to ``break the usual routine''~\cite{Mesquida2016}.
They help alleviate common Retro headaches by encouraging the exploration of new perspectives, increasing motivation and supporting team-building efforts~\cite{Mesquida2016}.

\section{Research Goals}
Explicit attributions of remedies to corresponding headaches currently only exist in a few cases, which have been identified as valuable tools for practitioners~\cite{jovanovic2016}.
In previous work, we proposed a mapping of Retro problems to activities, which can be employed by Agile practitioners to find the appropriate remedy for an identified headache~\cite{eurospi19}.
In this paper, we focus on the empirical evidence that underpins these connections.
The following research question (RQ) guides our work:

\boxbox{0.9}{\textbf{RQ:} Which activities employed by Agile process facilitators remedy identified headaches in Retrospective meetings?}

We present initial observational case studies on the effectiveness and feasibility of specific Retro remedies concerning the Retro headaches observed in Scrum software development teams, both in educational as well as professional contexts.
Our results indicate that while responses to the introduction of Retro activities were generally positive, the efficacy of alleviating the identified headache varied between different employed remedies.
Furthermore, we note that administering Retro remedies in teams was also connected to the development of adverse effects in some observed scenarios, such as additional Retro headaches.
These results highlight the importance of selecting the appropriate Retro activity as a remedy for an identified headache as well as the need for detailed information on remedies, i.e. a ``medication package insert'', for Agile practitioners.

\section{Related Work}
Activities, which offer team actions for meetings and keep them fresh have been used in Retros since their inception.
In 2000, Kerth initially published a collection of activities for Retros~\cite{kerth2000}.
In the following years, collections were published by multiple further authors~\cite{hohmann2006,kua2013,goncalves2014,krivitsky2015}.
Derby and Larsen introduced a general agenda for Retros, featuring five consecutive phases---set the stage, gather data, generate insights, decide what to do, close---and proposed activities for each stage~\cite{Derby2006}.
This concept was adopted by other authors~\cite{Baldauf2018}.
Jovanovic et al.~\cite{jovanovic2016} mapped activities to the four phases of Tuckman's teamwork model, i.e. forming, storming, norming and performing~\cite{tuckman1965}.
However, only a few authors, such as Kua~\cite{kua2013} and Rubin~\cite{rubin12}, have connected the proposed activities to specific headaches that have been observed.
They suggest, e.g. to ask the team targeted questions at opportune moments.

Esther and Larsen propose to ``start with the hard data'' at the beginning of a Retro~\cite{Derby2006}, collecting the teams' events, metrics and artifacts of the last development iteration.
This approach supports governing and controlling the executed development process based on empirical evidence~\cite{loeffler2017}.
This focus on empiricism, i.e. basing decisions on knowledge derived from experience and project data, is reflected in the theory of the Scrum framework, with the concept of ``empirical process control''~\cite{Schwaber2017}.
However, these findings from theory and research have often not reached standard industry practice.
In a study of Microsoft developers, Devanbu et al. found that programmers beliefs, rather than being founded on empirical insights, were primarily formed based on personal experience~\cite{Devanbu2016}.
The way that Retros are run and possible headaches are tackled is then often based on anecdotal evidence and personal perceptions.
Evidence-based medicine, with its focus on disseminating empirical research results to practitioners, can serve as a model to change this.
By producing and publishing more empirical results of real Retro activity usage in case studies or experiments, researchers and practitioners alike can benefit from evidence-based practice.

\section{Method}
\label{sec:method}
The goal of this research is to provide additional empirical evidence for the effectiveness of specific Retro activities in remedying commonly identified headaches in the context of the Scrum method.
We aim at supporting the roles tasked with facilitating and improving the software development process in teams in running more effective and enjoyable Retros.

We performed the following steps towards this goal and to answer our previously defined research question:
\begin{enumerate}
    \item Collect Retro activities and the headaches they aim to solve from research literature and previous work.
    \item Conduct real-world case studies in Scrum teams, advising process facilitators and teams to introduce appropriate Retro remedies for identified headaches.
    \item Collect perceptions of team members after remedy application using surveys. Evaluate the influence of activities on the treated headaches.
\end{enumerate}

\noindent The following subsections detail these individual steps.

\subsection{Retrospective Activities \& Headaches in Literature}
\label{sec:activity_problem_sources}
Work by Jovanovi{\'{c}} et al.~\cite{jovanovic2016} and Loeffler~\cite{loeffler2017} as well as our own previous work~\cite{eurospi19} includes structured collection efforts of both Retro headaches and activities.
We unified and deduplicated the employed primary sources, extracting descriptions and instructions of activities as well as headaches.
We also explicitly included up-to-date web resources as part of our sources, as these are used by Agile practitioners on a daily basis~\cite{loeffler2017,Beecham2014}, in contrast to published research~\cite{Devanbu2016}.

\Cref{table:activity_src} lists the primary sources of Retro activities and their descriptions, which were used as possible remedies in the case study.
Common Retro headaches to be looked out for in meetings of the case study were extracted from the sources listed in \Cref{table:problems_src}.

\begin{table}[htb]
    \centering
    \caption{Sources of Retrospective activities.}
    \label{table:activity_src}
    \begin{tabularx}{\columnwidth}{@{}lX@{}}
        \toprule
        \textbf{Year} & \textbf{Name and Reference}\\
        \midrule
        2006 & Agile Retrospectives - Making Good Teams Great~\cite{Derby2006} \\
        2006 & Innovation Games: Creating Breakthrough Products through Collaborative Play~\cite{hohmann2006} \\
        2013 & The Retrospective Handbook~\cite{kua2013} \\
        2013 & Project Retrospectives: A Handbook for Team Reviews~\cite{kerth2013} \\
        2014 & Getting value out of Agile Retrospectives: A Toolbox of Retrospective Exercises~\cite{goncalves2014} \\
        2015 & Agile Retrospective Kickstarter~\cite{krivitsky2015} \\
        2015 & Fun Retrospectives - Activities and ideas for making agile retrospectives more engaging~\cite{caroli2015} \\
        2018 & Retromat: Run great agile retrospectives!~\cite{Baldauf2018} \\
        2019 & Partnerships and Possibilities Blog~\cite{Larsen2019} \\
        2019 & Agile Retrospective Resource Wiki: Retrospective Plans~\cite{RetroWiki2019Plans} \\
        \bottomrule
    \end{tabularx}
\end{table}

\begin{table}[htb]
    \centering
    \caption{Sources of Retrospective headaches.}
    \label{table:problems_src}
    \begin{tabularx}{\columnwidth}{@{}lX@{}}
        \toprule
        \textbf{Year} & \textbf{Name and Reference}\\
        \midrule
        2012 & Essential Scrum~\cite{rubin12} \\
        2013 & The Retrospective Handbook~\cite{kua2013} \\
        2014 & Do's and Don'ts of Agile Retrospectives~\cite{Amin14} \\
        2017 & Improving Agile Retrospectives~\cite{loeffler2017} \\
        2017 & 9 Deadly Agile Retrospectives Antipatterns~\cite{Goncalves2017} \\
        2018 & Agile Retrospective Wiki: Common Ailments and Cures~\cite{Bowley2018} \\
        \bottomrule
    \end{tabularx}
\end{table}

\subsection{Observing Headaches and Administering Remedies}
To gain an understanding of the Retro headaches and the remedies that could be employed, we conducted multiple case studies~\cite{flick2009}.
We took part in meetings in the role of ``observer as participant''~\cite{Gold1958}. Team members were aware of the presence of the researcher and were previously informed of the goals of the research.

Overall, we were present in nineteen Retros as part of the performed case studies. The Retros were held in three distinct contexts: an undergraduate Agile software development university course, a mobile gaming startup and a large established software development company.
The university course was a capstone project course in the undergraduate program, focused on the application of Scrum in a practical software development project, involving the collaboration of four self-organizing teams~\cite{Matthies2016c}. In the rest of this paper, we refer to the four university Scrum teams as Teams A, B, C, and D. Thirty-one students in the final year of their undergraduate studies with experience in programming as well as scrum participated in the course. The course project was split into four Sprints, each two weeks long, and included Daily Scrums, Sprint Plannings, Sprint Reviews, as well as Retros organized by the participants. 
While results obtained from student teams are not immediately generalizable to other contexts, related work also points out that performance between professionals and students did not differ when experiments involved applying a new technology~\cite{Salman2015}.
The mobile gaming startup team worked in two-week Sprints and teams conducted short Retros after the end of each iteration, often in the form of ``I like, I wish sessions''~\cite{Both2014}.
The team from an established global player in the software industry consisted of an entire department with three teams of eight developers, three designers, a part-time Scrum Master, a Product Owner, and a manager. In addition to the Retros performed within the scrum teams, quarterly Retros with the whole department were performed.

Five Retros were solely observed, to gain an initial understanding of the teams' contexts.
In fourteen Retros we proposed activities to the Scrum Master, based on previously identified Retro headaches, influencing the way the meetings were organized and run.
In these cases, we proposed specific activities to the Scrum Masters of observed teams, who then implemented them.
The headaches to be addressed were identified through structured interviews with the team's Scrum Master or, where possible, by analyzing the notes of the previously observed Retro.
Activities to remedy the identified headaches were selected and proposed to the Scrum Masters in individual one-on-one meetings in which the selected activity, its structure, and goals, were thoroughly explained.
The mapping that was used to match Retro headaches to activities that remedy them is detailed in \Cref{sec:mapping}.
If multiple headaches were noticed in a single Retro, mappings were searched for activities that could address both headaches simultaneously.
If these could not be found, two separate appropriate activities were employed consecutively in teams.

\subsection{Collecting Perceptions of Interventions}
We applied three data collection methods to capture the perceptions of team members towards the employed interventions and their effects:
(i) We observed the interventions and took notes.
(ii) We asked team members of facilitated Retros to fill out surveys at the end of each meeting.
(iii) We interviewed the six Scrum Masters of facilitated Retros after their final Retros of the case study.

The employed surveys captured the perceptions of team members towards the used activities and their effects with a mixture of closed and open format items as defined by Leung~\cite{leung2001}.
Accordingly, every survey contained an item capturing the enjoyment of participants regarding the applied Retro activity, e.g. ``Did you like the Sailboat activity?''.
The item could be answered with \emph{Yes}, \emph{No} or \emph{Other}.
When selecting the Other option, participants were encouraged to add explanatory free text.
Additionally, the surveys contained two free text questions asking for positive and negative aspects of the Retro and an optional comment field for feedback.

The final interviews with Scrum Masters were concerned with their impressions of the activities as well as success in remedying the identified headaches.
As the Scrum Masters were tasked with implementing the proposed activities and possess intricate knowledge of their teams they represent an important source of knowledge and insights.
Interviews were conducted at the very end of the case studies with the industry partners as well as with the students.
The interviews were recorded and transcribed.
We analyzed the collected data from observations, open survey items, and interviews through coding~\cite{flick2009}.
Notes, participant answers, and interview transcripts were iteratively labeled with the perceived effects of each activity towards problematic aspects of Retros.
The closed format question was mapped to a linear scale of -1 (\emph{No}), 0 (\emph{Other}) and 1 (\emph{Yes}).
We calculated averages based on all given answers and this mapping.
The results of these steps are presented in \Cref{fig:radars}.

\section{Employed Headache-Remedy Mappings}
\label{sec:mapping}
To propose the appropriate remedy for an identified Retro headache, a mapping between them has to be established.
In this case study, we observed five common Retro headaches in the meetings of participating Scrum teams, as listed in \Cref{table:retro_problems}. 
\begin{table*}[htb]
    \caption{Retrospective headaches and how often they were observed and addressed in the case studies.}
    \label{table:retro_problems}
    \centering
    \begin{tabularx}{\textwidth}{lXr}
        \toprule
        \textbf{Headache}  & \textbf{Definition} & \ Qty   \\
        \midrule
        \multicell{No Preparation \cite{rubin12,kua2013,loeffler2017}} & Few arrangements by facilitator, participants' time not being valued, lack of structure. & 5\\
        \arrayrulecolor{gray} \cmidrule(lr){1-3} \arrayrulecolor{black}%
        \multicell{Not Speaking Up \cite{rubin12,Amin14,Bowley2018}} & Reluctance of team members to reflect or to share perceptions of (known) problems. & 4\\
        \arrayrulecolor{gray} \cmidrule(lr){1-3} \arrayrulecolor{black}%
        \multicell{All Talk--No Action \cite{kua2013,loeffler2017,Goncalves2017,Bowley2018}} & Few outcomes defining the next improvement steps, no clear path for improvement. & 3\\
        \arrayrulecolor{gray} \cmidrule(lr){1-3} \arrayrulecolor{black}%
        \multicell{Focus on Negatives \cite{loeffler2017,Bowley2018}} & Positive aspects of previous iteration disregarded in favor of negatives, leading to low team spirit. & 1\\
        \arrayrulecolor{gray} \cmidrule(lr){1-3} \arrayrulecolor{black}%
        \multicell{Too Repetitive \cite{kua2013}} & Unchanging Retro procedures, leading to fatigue, frustration and low motivation. & 1\\
        \bottomrule
    \end{tabularx}
\end{table*}
In a following step, Retro activities were defined or selected from related work.
These should be able to address the headaches and should apply to the teams' contexts and circumstances.
This mapping was drawn from previous work~\cite{eurospi19} and is presented in \Cref{table:mapping}.
It is based on the analysis of the detailed activity descriptions in the primary sources, cf. \Cref{sec:activity_problem_sources}.
It is this mapping that we aim to validate empirically.

\begin{table*}[htb]
    \caption{Mapping of observed Retrospective headaches to possible remedies, based on~\cite{eurospi19}.}
    \label{table:mapping}
    \centering
    \begin{tabularx}{\textwidth}{lX}
        \toprule
        \textbf{Headache}  & \textbf{Possible Remedies} \\
        \midrule
        No Preparation & \textbullet~Sailboat \textbullet~Futurespective \textbullet~Emotional Seismograph \textbullet~Open the Box \textbullet~Tweet My Sprint \\
        \arrayrulecolor{gray} \cmidrule(lr){1-2} \arrayrulecolor{black}%
        Not Speaking Up & \textbullet~Sailboat \textbullet~Emotional Seismograph \textbullet~Open the Box \textbullet~Story Oscars \textbullet~Tweet My Sprint \textbullet~Collective Painting \textbullet~Team Member Oscars \\
        \arrayrulecolor{gray} \cmidrule(lr){1-2} \arrayrulecolor{black}%
        Focus on Negatives & \multicell{\textbullet~Sailboat \textbullet~Futurespective \textbullet~Open the Box \textbullet~Story Oscars \textbullet~Circles and Soup\\\textbullet~Team Member Oscars} \\
        \arrayrulecolor{gray} \cmidrule(lr){1-2} \arrayrulecolor{black}%
        All Talk--No Action & \textbullet~Open the Box \textbullet~Reverse Brainstorming \\
        \arrayrulecolor{gray} \cmidrule(lr){1-2} \arrayrulecolor{black}%
        Too Repetitive & \multicell{\textbullet~Futurespective \textbullet~Circles and Soup \textbullet~Tweet My Sprint \textbullet~Collective Painting\\\textbullet~Reverse Brainstorming \textbullet~Guess Who \& Emotional Seismograph} \\
        \bottomrule
    \end{tabularx}
\end{table*}

\subsection{Retrospective Activity Details}
We summarize the eleven Retro activities employed in the case studies in the following paragraphs.
Activities marked with an asterisk (*) represent adaptations to existing activities made during the observational case study.
These were necessary due to the individual contexts of teams and were agreed on with Scrum Masters and process facilitators of teams.

\begin{itemize}[leftmargin=*]
    \item \emph{Sailboat}~\cite{goncalves2014}: Reflection on the past Sprint using a sailboat visualization including positives (wind), negatives (anchors), opportunities (sun) and dangers (icebergs) in the future. Team members share and categorize items after brainstorming.
    \item \emph{Open the Box}~\cite{caroli2015}: Brainstorming using a box metaphor. The box contains the team's actions of the last Sprint. The team decides which actions to add or to remove and which should stay.
    \item \emph{Futurespective}~\cite{hohmann2006}: The team imagines they are already conducting the next Sprint's Retro and brainstorms what made the current Sprint a success.
    \item \emph{Emotional Seismograph}~\cite{caroli2015}: Visualization activity where team members plot the development of their mood throughout the last sprint in a shared graph, with the axes representing time and happiness, also referred to as \emph{Peaks and Valleys Timeline}.
    \item \emph{Circles and Soup}~\cite{Baldauf2018}: Suggested improvements are divided into three areas based on the level of control team members can exert: (i) addressable by team members themselves, (ii) people outside the team are needed or (iii) the team has no influence (the soup).
    \item \emph{Story Oscars}~\cite{Baldauf2018}: Team members reflect on last Sprint's user stories. Stories are nominated for different award categories, e.g. most challenging. The team votes to decide winners.
    \item \emph{Tweet My Sprint}~\cite{Baldauf2018}: Team members write short informal notes/tweets concerning the past iteration, which are shared and arranged in a timeline, and can be ``retweeted'' or replied to.
    \item \emph{Guess Who}~\cite{Baldauf2018} \& Emotional Seismograph*: The Emotional Seismograph activity is performed by individual team members. The others guess which mood graph belongs to which team member, stating their rationales.
    \item \emph{Reverse Brainstorming*}: The team brainstorms using the prompt ``What would make the next Sprint the worst ever?''. Then the next steps can be defined based on the logical opposite of items.
    \item \emph{Team Member Oscars*}: Adaptation of the Story Oscars activity. Team members award each other based on work of the last Sprint. Every participant should receive an award.
    \item \emph{Collective Painting*}: Free-form activity, where the team, in silence, collectively paints a shared picture representing their perceptions of the past sprint.
\end{itemize}

\section{Case Study Results}
\begin{figure}[htb]
    \centering
	\includegraphics[width=0.95\linewidth]{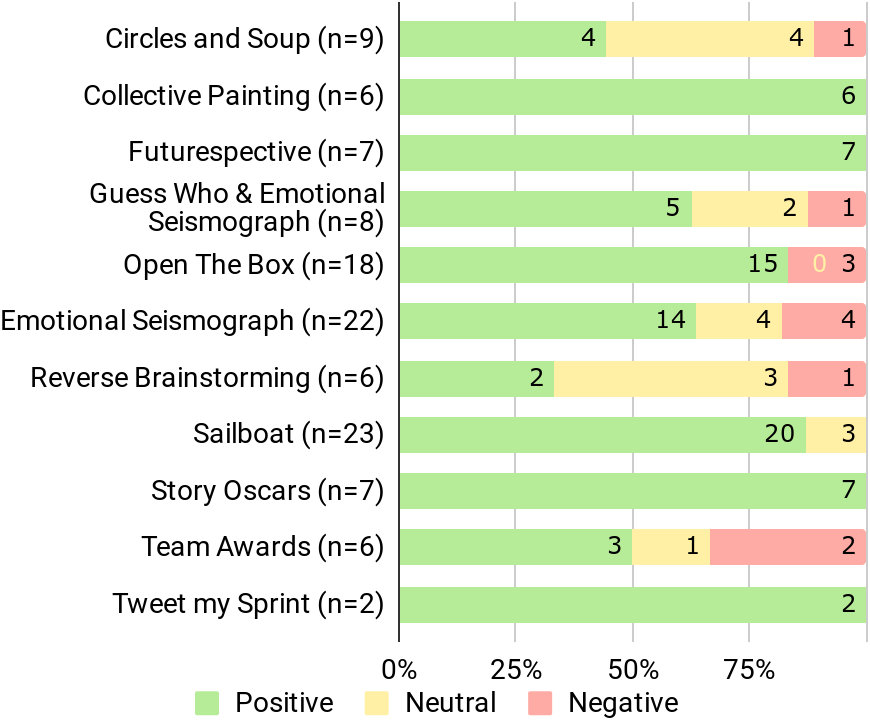}
	\caption{Mean perceptions of teams regarding the introduction of Retrospective activities.}
	\label{fig:general_rating}
\end{figure}

During the Retros in our case study, we observed five previously described headaches, cf. \Cref{table:retro_problems}.
The most common headaches in our case study were \emph{No Preparation}, \emph{Not Speaking Up} and \emph{All Talk--No Action}.
\emph{Too Repetitive} could only be observed once. In all of these cases, the headaches lead to serious problems during the Retros such as meetings running much longer than planned, finishing without clear action items to improve the next sprint or not discussing problems that existed during the sprint. \emph{Focus on Negatives} did not lead to problems within the Retros of \emph{Team A}, however, the team tended to be negative so we included it in the problem list and decided to remedy this headache with an activity in \emph{Team A R4}.  
Due to the nature of this observational case study, we relied on headaches naturally occurring during observations. 
Therefore, we can only offer evidence for the described set of headaches.
Scrum Masters and Agile facilitators participating in the case study understood the proposed Retro activities and were able to apply them in their teams in all observed cases. With their help, eleven separate activities were employed to remedy identified headaches in subsequent meetings.
\Cref{fig:general_rating} presents the activities as well as the results of surveys regarding the mean perceptions of activity introduction by team members.
Overall, we collected 122 filled out questionnaires for 18 interventions conducted within 14 Retros.
Out of the 11 activities 8 received positive or mostly positive ratings. Team Awards received the most negative ratings from all activities.

\Cref{table:case_study_retros} presents the details of all Retros that we facilitated as part of the case studies, along with the headache that was identified and the specific remedy that was applied.
\begin{table*}[htb]
    \centering
    \begin{tabularx}{\textwidth}{@{}lllXl}
        \toprule
        \textbf{Retrospective} & \textbf{Headache} & \textbf{Remedy} & \textbf{Results}\\
        \midrule
        Team A R2 & \multicell{No Preparation,\\Not Speaking Up,\\(Focus on Negatives)} & Sailboat & \cellcolor{info}\multicell{Headaches remedied, but new headache arose}\\
        Team A R3 & \multicell{All Talk--No Action,\\ (Focus on Negatives)} & Reverse Brainstorming & \cellcolor{error} Small improvements, but headache persists \\
        Team A R4 & Focus on Negatives & Team Member Oscars & \cellcolor{success} Headache remedied \\
        & All Talk--No Action & Open the Box & \cellcolor{success} Headache remedied \\
        \arrayrulecolor{gray} \cmidrule(lr){1-4} \arrayrulecolor{black}%
        Team B R2 & No Preparation & Futurespective & \cellcolor{success} Headache remedied \\
        Team B R3 & - & Emotional Seismograph & \cellcolor{success}\multicell{Variety recognized: more emotional topics} \\
        Team B R4 & - & Circles and Soup & \cellcolor{warning}\multicell{Variety recognized, method hard to understand} \\
        & - & Story Oscars & \cellcolor{warning}\multicell{Variety recognized, method took too long} \\
        \arrayrulecolor{gray} \cmidrule(lr){1-4} \arrayrulecolor{black}%
        Team C R2 & No Preparation & Emotional Seismograph & \cellcolor{success} Headache remedied \\
        Team C R3 & - & Sailboat & \cellcolor{success} Variety recognized \\
        Team C R4 & - & \multicell{Guess Who +\\Emotional Seismograph} & \cellcolor{warning}\multicell{Variety recognized, but few action items defined,\\might lead to All Talk--No Action} \\
        \arrayrulecolor{gray} \cmidrule(lr){1-4} \arrayrulecolor{black}%
        Team D R2 & - & Sailboat & \cellcolor{success} Variety recognized \\
        Team D R3 & - & Collective Painting & \cellcolor{info}\multicell{Variety recognized, but new headache arose}\\
        Team D R4 & Not Speaking Up & Tweet My Sprint & \cellcolor{success} Headache remedied \\
        & - & Circles and Soup & \cellcolor{warning}\multicell{Variety recognized, method hard to understand}\\
        \arrayrulecolor{gray} \cmidrule(lr){1-4} \arrayrulecolor{black}%
        Startup I R1 & \multicell{Not Speaking Up} & Emotional Seismograph & \cellcolor{success} Headache remedied\\
        & \multicell{Too repetitive,\\No Preparation} & Sailboat & \cellcolor{info} \multicell{Headaches remedied, but new headache arose}\\
        Global Player & No Preparation & Open the Box & \cellcolor{success} Headache remedied\\
        \bottomrule
    \end{tabularx}
    \caption{Details of case study Retrospectives. Remedy applications are marked as successful (green), generating adverse effects (blue), ambiguous (yellow) and unsuccessful (red). Parenthesis denote latent headaches, ``-'' denotes usage of activities for feedback collection.}
    \label{table:case_study_retros}
\end{table*}
%
Of the 18 instances in which activities were employed in Retros, 8 activities did not address a specific problem and were introduced for variety and to collect perceptions about the method from the teams.
Of the 10 activities that were employed to remedy a headache, 9 cases were successful.
Only once (\emph{Team A R3}, Reverse Brainstorming) did the identified headache persist after the team employed the corresponding activity.
However, in three cases the introduction of a Retro activity led to new headaches, which had previously not been observed in the team.

\begin{figure}[htb]
    \centering
	\includegraphics[width=0.95\linewidth]{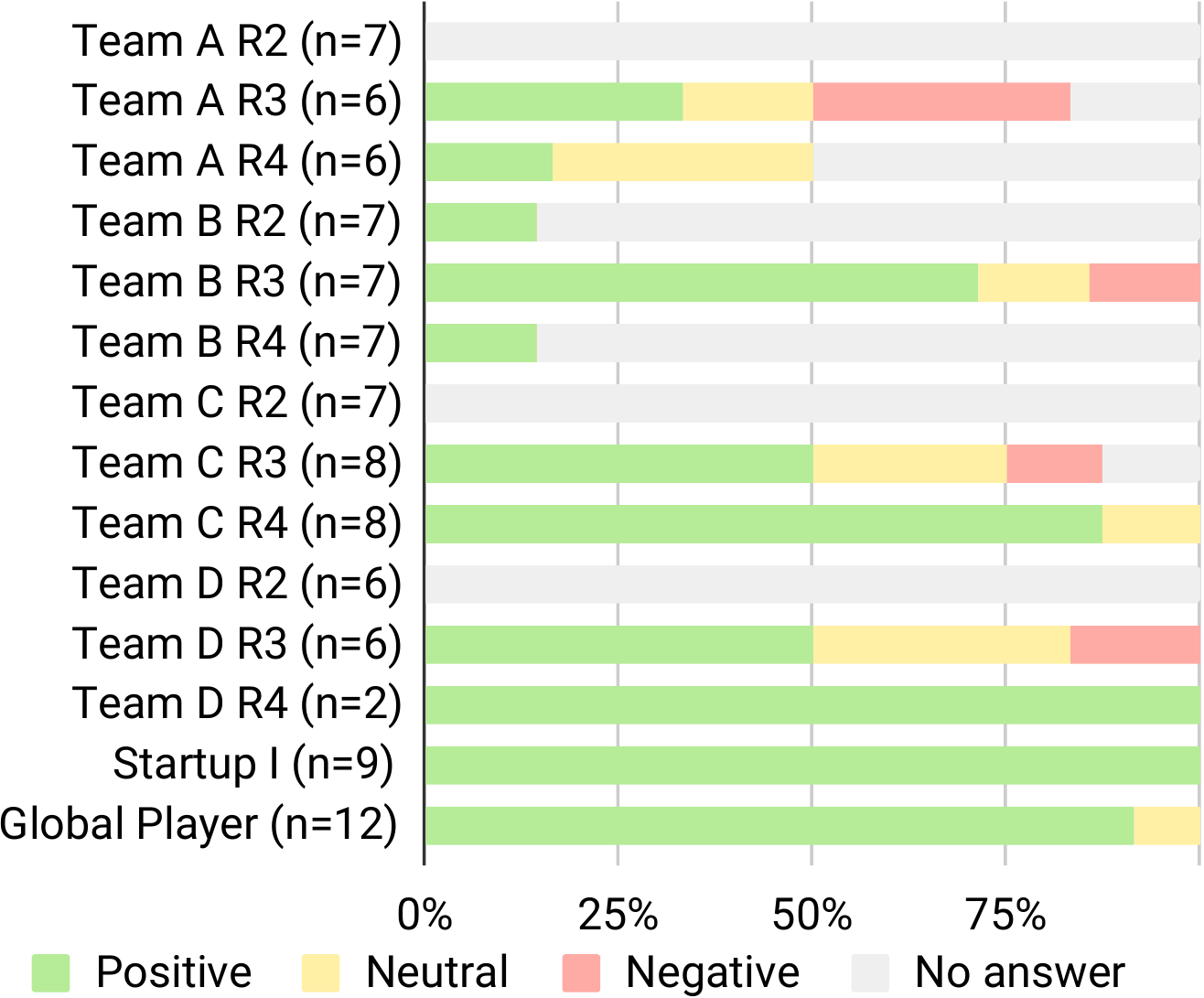}
	\caption{Mean perceptions of team morale improvement after each Retrospective.}
	\label{fig:team_spirit}
\end{figure}

We did not track the effect of our Retros in terms of team velocity.
However, we asked participants to rate their team morale after each Retro, see \Cref{fig:team_spirit}.
The number of answers to this question varied largely for the student teams.
There were few negative effects reported and 7 of the 11 retros were associated with improvement in team morale.

\begin{figure*}[htb]
    \centering
	\includegraphics[width=0.3\linewidth]{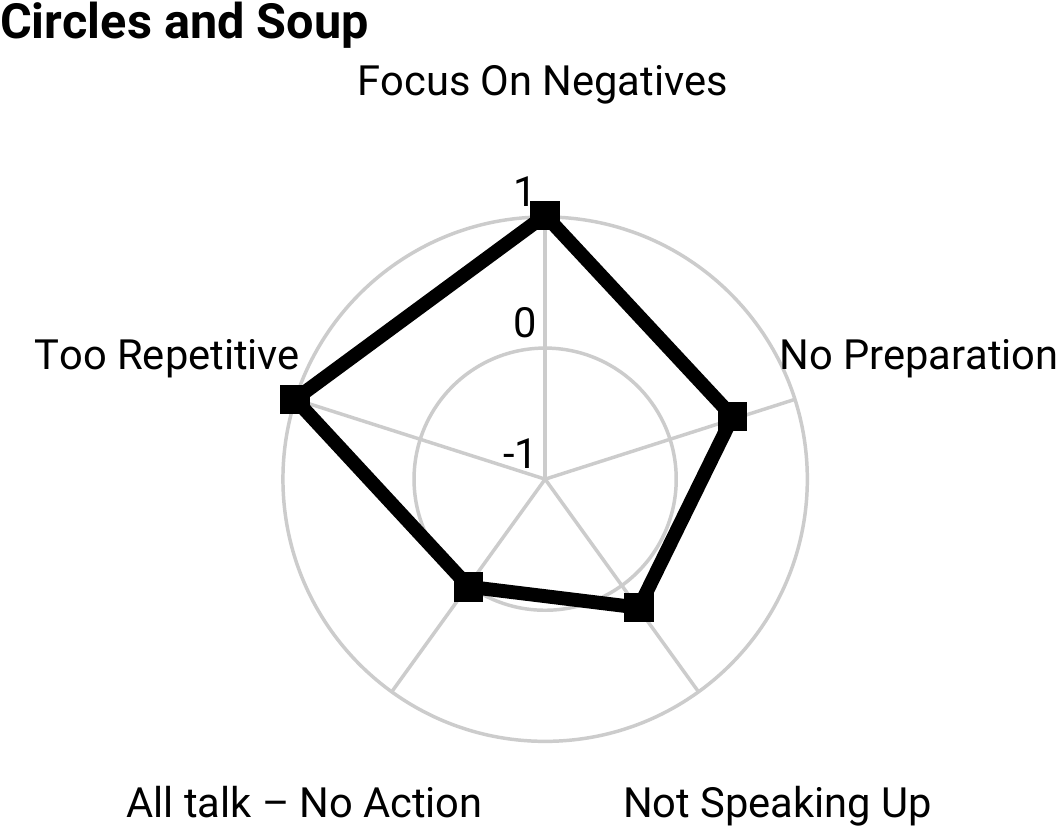}
	\includegraphics[width=0.3\linewidth]{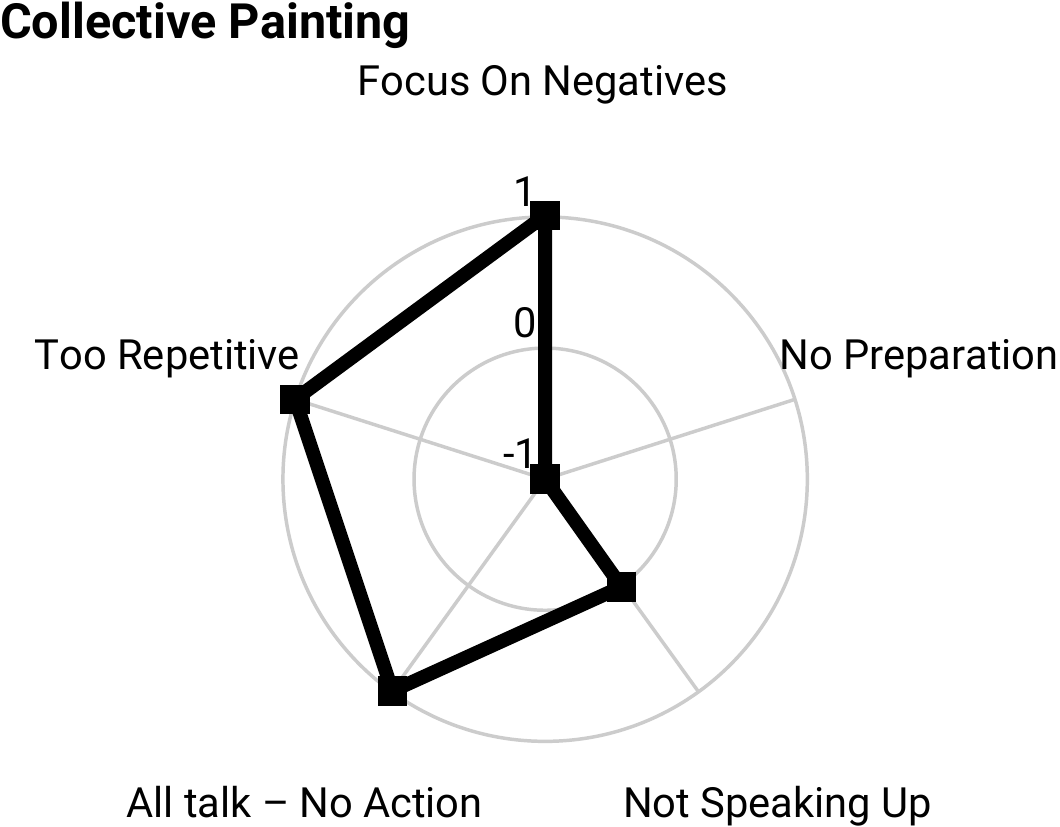}
    \includegraphics[width=0.3\linewidth]{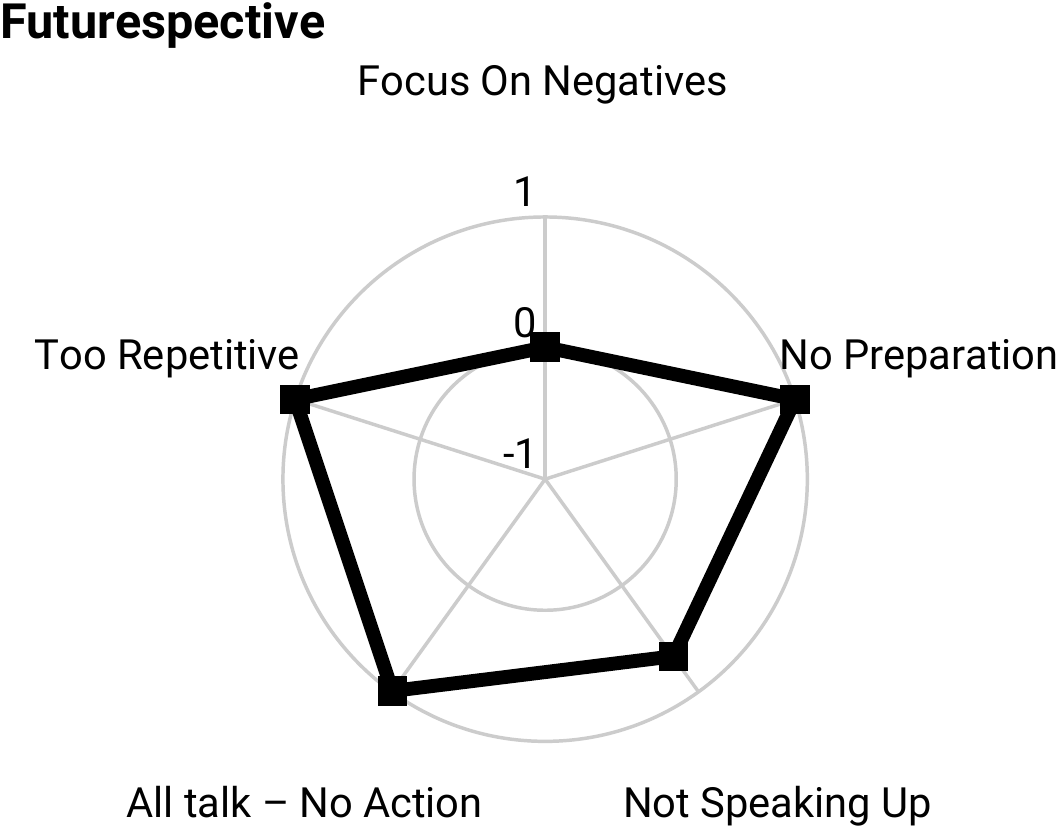}
    
    \vspace{1em}
    
    \includegraphics[width=0.3\linewidth]{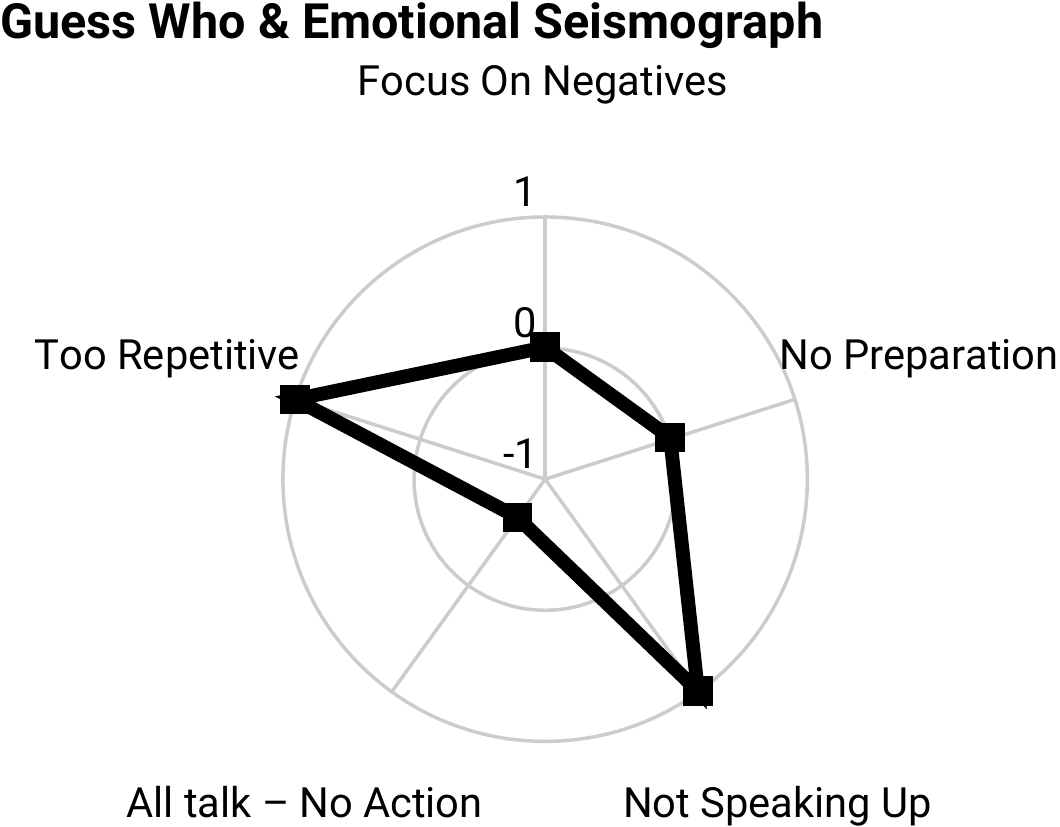}
    \includegraphics[width=0.3\linewidth]{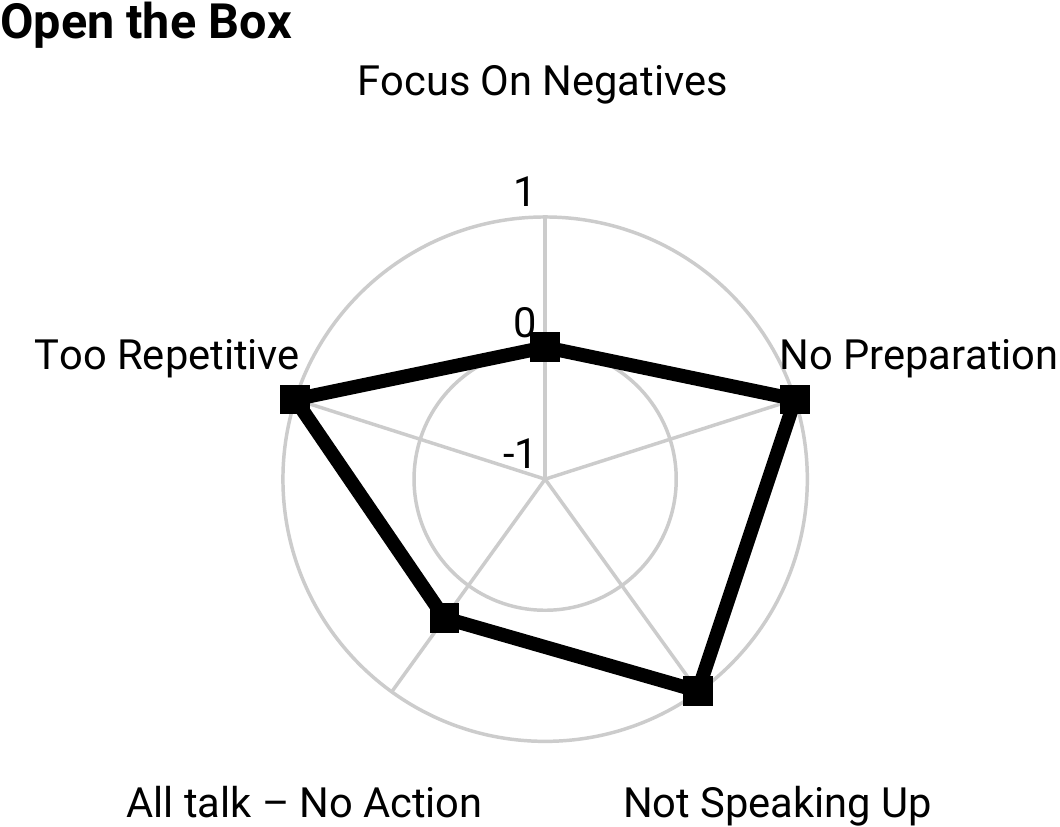}
    \includegraphics[width=0.3\linewidth]{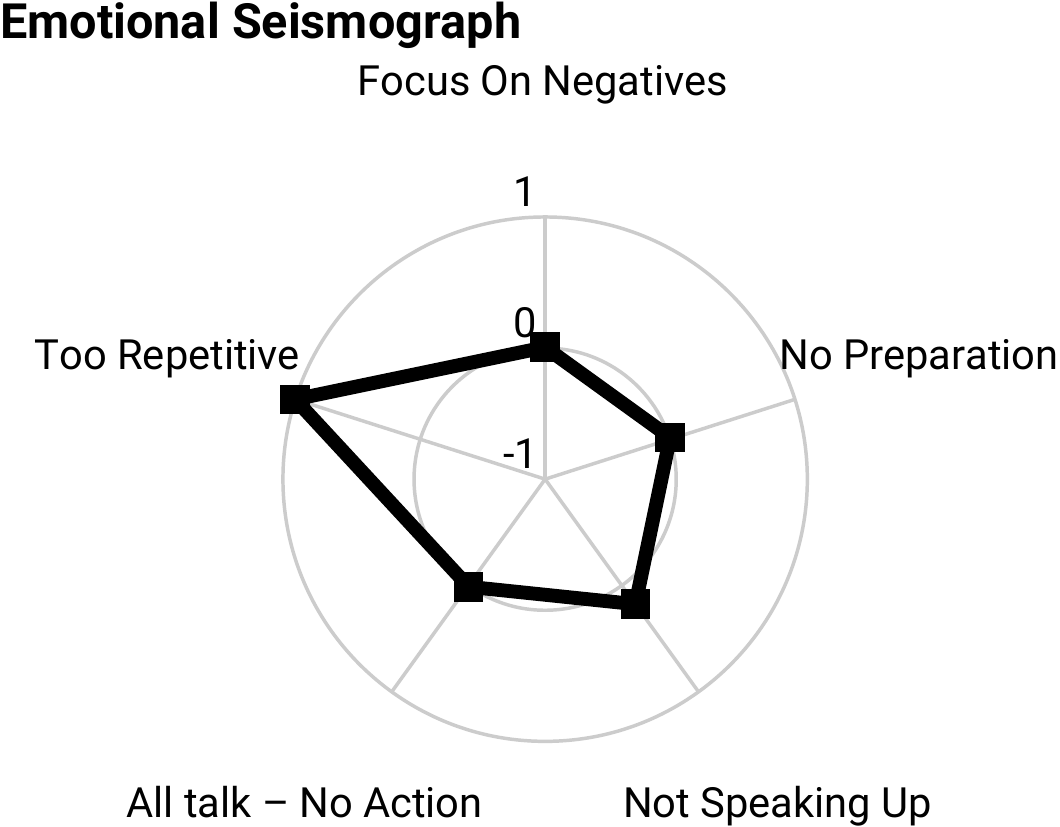}
    
    \vspace{1em}
    
    \includegraphics[width=0.3\linewidth]{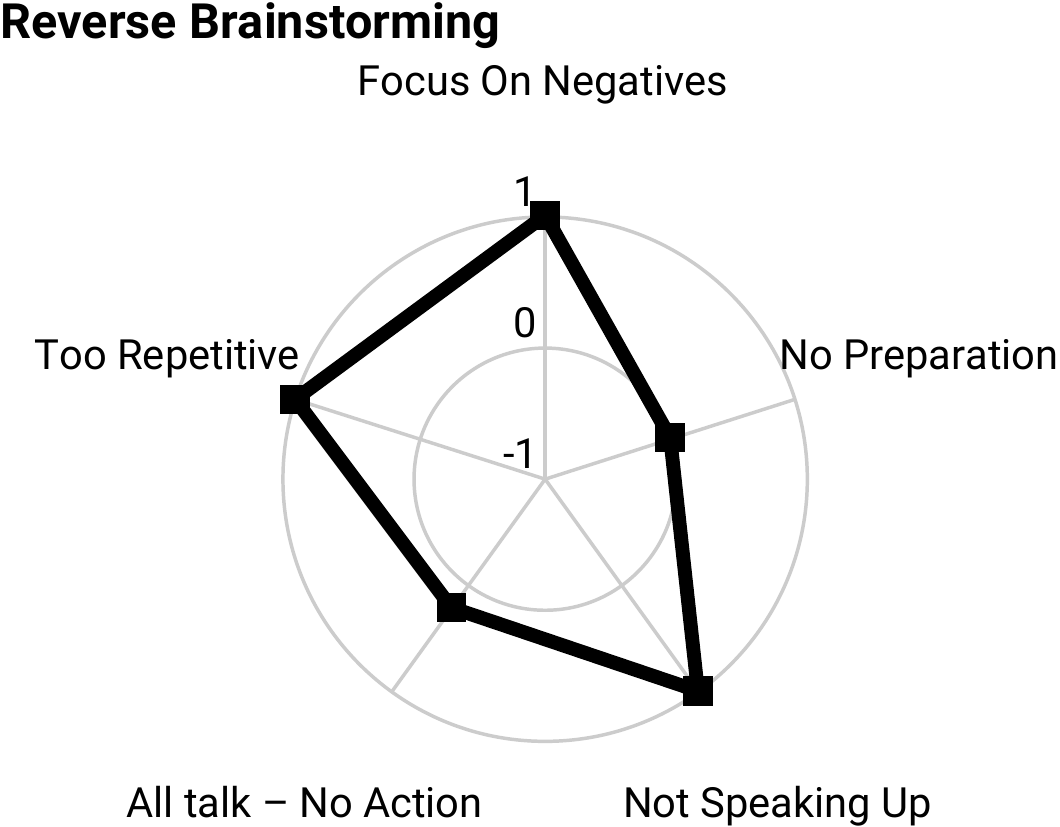}
    \includegraphics[width=0.3\linewidth]{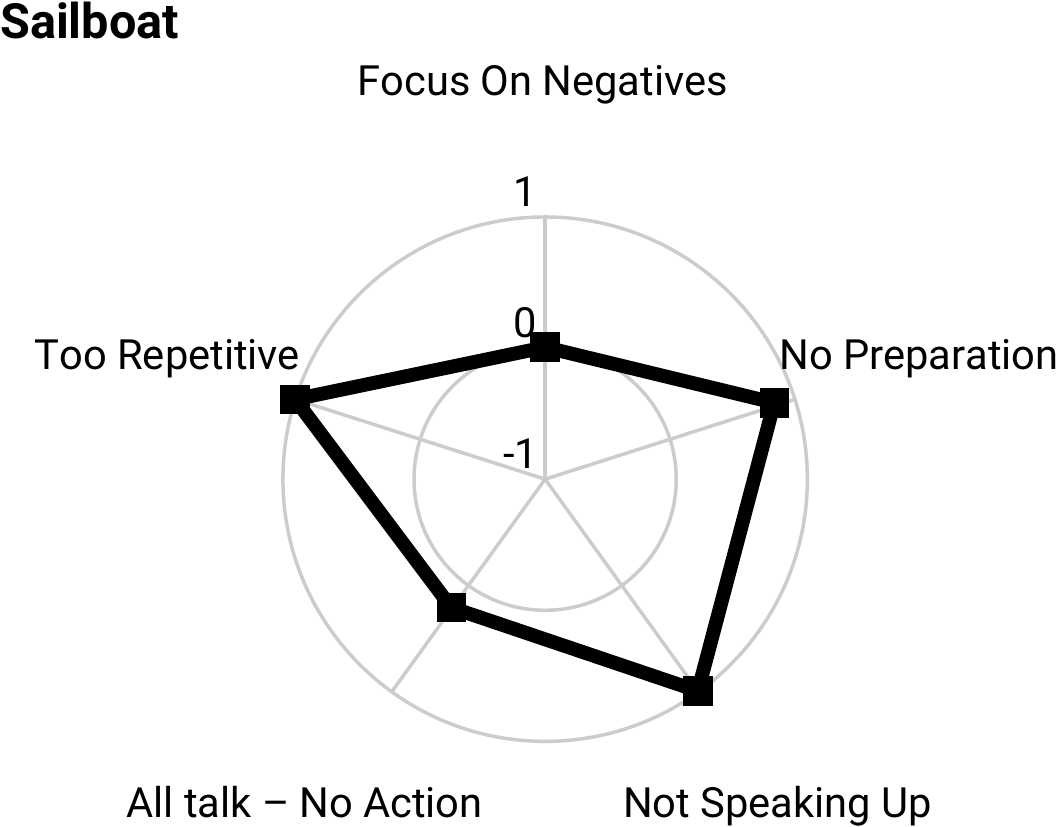}
    \includegraphics[width=0.3\linewidth]{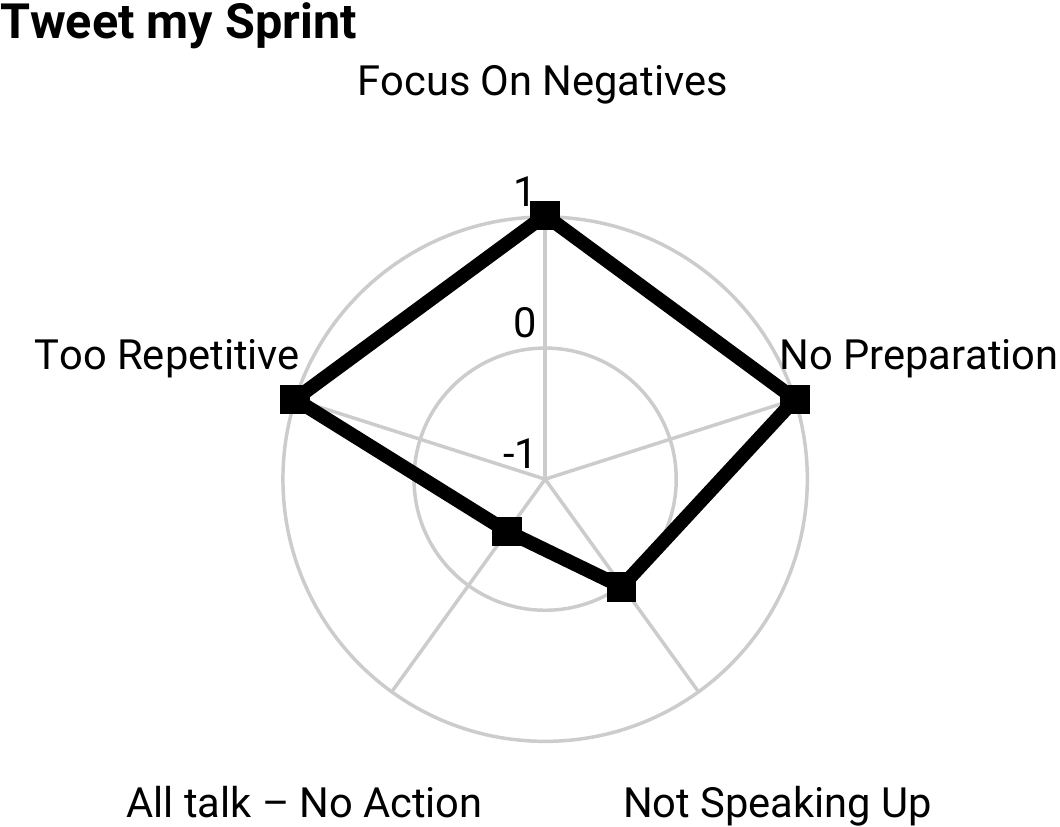}
	\caption{Radar charts of the perceived mean influence (0: neutral, -1: negative, 1: positive) of different Retrospective activities on meeting headaches.}
\label{fig:radars}
\end{figure*}

We analyzed our collected observation notes, free text survey answers and interview transcripts for mentions of positive and negative influences on problematic aspects of Retros.
For example, one participant explained that Reverse Brainstorming helped him ``to realize what has worked well'', which we noted as an indicator, that this method can help to address Focus On Negatives.
Another participant stated that the Sailboat ``provides a good structure along which we could discuss the different topics'' which we took as an indicator that the method could address No Preparation.
Regarding Circles and Soup, a participant mentioned ``it was a completely different way of doing things and thus provided a totally new point of view'' which we took as an indicator that this method addresses Too Repetitive.
Based on these mentions, we created profiles for all the activities, that indicate team members' perceptions regarding the effectiveness of Retro activities in addressing identified headaches.
\Cref{fig:radars} presents these profiles in the form of radar charts.
The higher the value for a given dimension is, the more participants believed that the activity could improve an aspect of their Retros.
Enough data to construct a meaningful diagram was created through the coding steps for nine out of the eleven distinct activities that were part of the case studies.
The activities Story Oscars and Team Member Oscars received few mentions and ratings.
Case study participants positively mentioned a change or increased variety (dimension Too Repetitive) concerning all employed activities.
The Retro profiles indicate that the Futurespective was the best-suited \emph{allrounder} of all employed activities, being rated highly in all but one of the analyzed dimensions.
Circles and Soup, Open the Box, Reverse Brainstorming, and Sailboat, presented a mix of specific strengths and weaknesses.
While the Emotional Seismograph activity was able to successfully remedy headaches in teams, cf.~\Cref{table:case_study_retros}, there were strong discrepancies in participant ratings, leading to an average neutral rating.
The similar structure and steps of the activities Open the Box and Sailboat is reflected in their radar chart profiles.
Both activities make use of brainstorming and clustering, although different metaphors and categories are employed.

\section{Discussion}
The finding that participants, in general, enjoyed the proposed team activities is in line with related work on Retrospectives~\cite{Derby2006,jovanovic2016}.
Retro activities are designed to be engaging and satisfying.
However, employing them might lead to the team enjoying the activity and their shared time even though the original Retro goals are not accomplished.
The perceptions of team members regarding individual activities, cf. \Cref{fig:general_rating}, thus have to be interpreted with this idea in mind.
As Derby and Larsen point out, a guiding principle for Retros should be: ``have fun, but have a purpose''~\cite{Derby2006}.

As successful Retro activities rely on effective collaboration and shared team spirit, the experience of every individual team member is essential.
Participants who do not have a positive experience with a given activity can negatively impact the entirety of a Retro.
This is relevant for choosing activities based on our findings.
The presented perceptions gathered with our case studies can give initial clues for activity selection, i.e. preferring those with no or few negative perceptions, which can then be fine-tuned for a given context, cf. \Cref{fig:general_rating}.
Based on this rationale, we recommend using the Sailboat activity, which was used repeatedly and successfully in addressing multiple headaches (No preparation, Not speaking Up, Too Repetitive) and received no negative comments from the participants that employed it.
Similarly, we would recommend using activities other than Team Awards, as it received more than 25\% negative ratings.
Based on our findings, when confronted with a choice of multiple activities that address the same headaches, those with more positive perceptions should, therefore, be given priority.

Another indication for the importance of distinguishing activity enjoyment from activity success in addressing Retro headaches is the Reverse Brainstorming activity.
We considered the application of this technique in the observed Retro as not remedying the All Talk--No Action headache, cf. \Cref{table:case_study_retros}.
However, while half of the participants rated the activity as neutral, only less than a quarter rated it negatively.

In eight Retros we did not observe a pressing headache that required immediate remedy.
In these cases, we employed different activities to keep meetings varied and avoid the headache of Too Repetitive.
All seven distinct activities that were employed, successfully provided meeting diversity.
However, in half of the cases, the introduction of a novel activity, led to variety, albeit limiting the effectiveness of the resulting meeting.
In the other half, the outcomes of the Retro were not negatively affected.
Based on these findings, usage of activities that did not show adverse effects, e.g. Emotional Seismograph, should be prioritized.
It is worth noting, that repetitiveness is a latent headache in most teams, which requires constant vigilance to remedy and prevent.
Even well-functioning teams with successful Retros can benefit from introducing fresh ideas into their meetings.

The headache which was most frequently observed in case study teams was No Preparation.
In all five instances, the employed corresponding Retro activity led to a remedy of the headache.
The preplanned agendas of activities provide inherent structure and reduce preparation effort of process facilitators~\cite{Derby2006}.
However, the level of offered structure, and therefore, likely also activity effectiveness, varies between individual activities.
While the Emotional Seismograph activity successfully remedied No Preparation, it was noted that other activities were better suited.

While the overwhelming majority of employed Retro activities led to the resolution of the identified headaches, three out of 18 instances (16.7\%) also led to new headaches.
These observations highlight the fact that while activities have many benefits, they are also associated with costs, in terms of commitment, time and risk.
This study, therefore, represents first steps in the direction of ``medication package inserts'' for the application of Retro remedies.

\section{Conclusion}
The literature on Scrum and Agile software development methods contains multiple proposals for using team activities in Retros.
However, it is often unclear in which circumstances these activities should be employed and which common headaches or challenges they address.
Little empirical evidence and few validations of existing mappings of headaches to remedies are available.
In this paper, we report on case studies in which eleven separate Retro activities were used to address five distinct observed headaches in software development teams.
We identified previously described headaches in six Scrum teams of both professional and educational backgrounds.
Our results show that the vast majority of interventions using appropriate Retro activities led to the remedy of the identified headache.
Thus, we provide empirical evidence for the applicability of a headache-remedy mapping.
Accordingly, our results point to the following answer regarding our initial research question (RQ):
\boxbox{0.95}{\textbf{RQ answer:} We present empirical evidence of eleven distinct activities remedying five common Retrospective headaches, see \Cref{table:case_study_retros}. If applicable, these activities should be preferred in practice.
If headaches are not covered by existing literature, we suggest using the presented radar charts of popular activities and their perceived influences on headaches, see~\Cref{fig:radars}, to select the most fitting activity.}

\subsection{Future Work}
Due to the nature of the conducted case studies, our initial empirical findings can only present evidence for a subset of observed headaches and applied activities.
Accordingly, future work should explore further Retro headaches and appropriate activities as well as revisit the headaches and activities presented here.
An additional set of experiments with more diverse groups and a randomized selection of activities can strengthen our initial results.
Furthermore, in this study, we measured the effectiveness of remedies in terms of effects within Retros.
Future work should investigate the effect of such activities on team productivity factors, e.g. by explicitly including team velocity or morale.
Finally, our results highlight the costs of activities, i.e. while Retros were overwhelmingly successful, some remedies had adverse effects.
This should be investigated in future work.

From a practical viewpoint, our results can help Agile practitioners to select appropriate activities, which have the highest chance of remedying an identified headache in their teams.
The presented radar charts, cf. \Cref{fig:radars}, with perceptions of case study participants regarding problematic Retro aspects, can be used to select the best fitting activity.
We have added to the empirical evidence regarding Retro activities in the research literature.
However, in line with related research~\cite{loeffler2017, Devanbu2016}, we also note that practitioners often rely on websites and frequently updated resources rather than research papers. Therefore, future work could focus on integrating new empirical findings from research into existing participatory web resources, such as wikis.

%% file: hicss51.bbl
\begin{thebibliography}{10}

\bibitem{StateOfAgile2019}
{CollabNet Inc.}, ``{13th Annual State of Agile Report},'' tech. rep., 2019.

\bibitem{ScrumAlliance2018}
{Scrum Alliance}, ``{State of Scrum 2017-2018: Scaling and Agile
  Transformation},'' tech. rep., 2018.

\bibitem{Matthies2018b}
C.~Matthies, R.~Teusner, and G.~Hesse, ``{Beyond Surveys: Analyzing Software
  Development Artifacts to Assess Teaching Efforts},'' in {\em 2018 IEEE
  Frontiers in Education Conference (FIE)}, pp.~1--9, IEEE, 2018.

\bibitem{dingsoyr2018}
T.~Dings{\o}yr, M.~Mikalsen, A.~Solem, and K.~Vestues, ``{Learning in the Large
  - An Exploratory Study of Retrospectives in Large-Scale Agile Development},''
  in {\em International Conference on Agile Software Development},
  pp.~191--198, Springer International Publishing, 2018.
\newblock DOI: \texttt{10.1007/978-3-319-91602-6{\_}13}.

\bibitem{Kniberg2015}
H.~Kniberg, {\em {Scrum and XP From the Trenches}}.
\newblock C4Media, 2nd~ed., 2015.

\bibitem{Schwaber2017}
K.~Schwaber and J.~Sutherland, ``{The Scrum Guide - The Definitive Guide to
  Scrum: The Rules of the Game},'' tech. rep., 2017.

\bibitem{Derby2006}
E.~Derby and D.~Larsen, {\em {Agile Retrospectives: Making Good Teams Great}}.
\newblock Pragmatic Bookshelf, 2006.

\bibitem{rubin12}
K.~S. Rubin, {\em {Essential Scrum: A practical guide to the most popular Agile
  process}}.
\newblock Addison-Wesley, 2012.

\bibitem{kua2013}
P.~Kua, {\em {The Retrospective Handbook: A guide for agile teams}}.
\newblock CreateSpace Independent Publishing Platform, 2013.

\bibitem{goncalves2014}
L.~Gon{\c{c}}alves and B.~Linders, {\em {Getting Value out of Agile
  Retrospectives - A Toolbox of Retrospective Exercises}}.
\newblock Lulu. com, 2014.

\bibitem{caroli2015}
P.~Caroli and T.~Caetano, ``{Fun Retrospectives --- Activities and ideas for
  making agile retrospectives more engaging},'' {\em Leanpub, Layton}, 2015.

\bibitem{Mesquida2016}
A.-L. Mesquida, M.~Jovanovic, and A.~Mas, ``{Process Improving by Playing:
  Implementing Best Practices through Business Games},'' in {\em Systems,
  Software and Services Process Improvement}, pp.~225--233, Springer, 2016.

\bibitem{jovanovic2016}
M.~Jovanovi{\'{c}}, A.-L. Mesquida, N.~Radakovi{\'{c}}, and A.~Mas, ``{Agile
  retrospective games for different team development phases},'' {\em Journal of
  Universal Computer Science}, vol.~22, no.~12, pp.~1489--1508, 2016.

\bibitem{eurospi19}
C.~Matthies, F.~Dobrigkeit, and A.~Ernst, ``{Counteracting Agile Retrospective
  Problems with Retrospective Activities},'' in {\em Systems, Software and
  Services Process Improvement} (A.~Walker, R.~V. O'Connor, and R.~Messnarz,
  eds.), (Cham), pp.~532--545, Springer International Publishing, 2019.
\newblock DOI: \texttt{10.1007/978-3-030-28005-5{\_}41}.

\bibitem{kerth2000}
N.~L. Kerth, ``{The ritual of retrospectives: how to maximize group learning by
  understanding past projects},'' {\em Software testing {\&} quality
  engineering}, vol.~2, no.~5, pp.~53--57, 2000.

\bibitem{hohmann2006}
L.~Hohmann, {\em {Innovation games: creating breakthrough products through
  collaborative play}}.
\newblock Pearson Education, 2006.

\bibitem{krivitsky2015}
A.~Krivitsky, {\em {Agile Retrospective Kickstarter}}.
\newblock Leanpub, Layton, 2015.

\bibitem{Baldauf2018}
C.~Baldauf, {\em {Retromat - Run great agile retrospectives!}}
\newblock Leanpub, Layton, 2018.

\bibitem{tuckman1965}
B.~W. Tuckman, ``{Developmental sequence in small groups.},'' {\em
  Psychological bulletin}, vol.~63, no.~6, p.~384, 1965.

\bibitem{loeffler2017}
M.~Loeffler, {\em {Improving Agile Retrospectives: Helping Teams Become More
  Efficient}}.
\newblock Addison-Wesley Professional, 2017.

\bibitem{Devanbu2016}
P.~Devanbu, T.~Zimmermann, and C.~Bird, ``{Belief {\&} evidence in empirical
  software engineering},'' in {\em Proceedings of the 38th International
  Conference on Software Engineering - ICSE '16}, (New York, New York, USA),
  pp.~108--119, ACM Press, 2016.

\bibitem{Beecham2014}
S.~Beecham, P.~O'Leary, S.~Baker, I.~Richardson, and J.~Noll, ``{Making
  Software Engineering Research Relevant},'' {\em Computer}, vol.~47, no.~4,
  pp.~80--83, 2014.

\bibitem{kerth2013}
N.~L. Kerth, {\em {Project Retrospectives: A Handbook for Team Reviews}}.
\newblock Dorset House eBooks, Pearson Education, 2013.

\bibitem{Larsen2019}
D.~Larsen and J.~Shore, ``{Partnerships and Possibilities Blog}.''
  \url{https://www.futureworksconsulting.com/blog/}, 2019.
\newblock [online] Accessed: 2019-01-16.

\bibitem{RetroWiki2019Plans}
{RetrospectiveWiki Contributors}, ``{Retrospective Plans}.''
  \url{http://retrospectivewiki.org/index.php?title=Retrospective_Plans}, 2019.
\newblock [online] Accessed: 2019-06-13.

\bibitem{Amin14}
H.~Amin, ``{Do's and Don'ts of Agile Retrospectives}.''
  \url{https://web.archive.org/web/20170711122247/https://www.scrumalliance.org/community/articles/2014/july/dos-and-don-ts-of-agile-retrospectives},
  2014.
\newblock [online] Accessed: 2018-01-02.

\bibitem{Goncalves2017}
L.~Gon{\c{c}}alves, ``{9 Deadly Agile Retrospectives Antipatterns Every Scrum
  Master Must Avoid}.''
  \url{https://luis-goncalves.com/agile-retrospectives-antipatterns/}, 2017.
\newblock [online] Accessed: 2019-01-31.

\bibitem{Bowley2018}
R.~Bowley and B.~Linders, ``{Common ailments {\&} cures}.''
  \url{http://retrospectivewiki.org/index.php?title=Common_ailments_%26_cures},
  2018.
\newblock [online] Accessed: 2019-01-11.

\bibitem{flick2009}
U.~Flick, {\em {An introduction to qualitative research}}.
\newblock Sage Publications Limited, 2009.

\bibitem{Gold1958}
R.~L. Gold, ``{Roles in Sociological Field Observations},'' {\em Social
  Forces}, 1958.

\bibitem{Matthies2016c}
C.~Matthies, T.~Kowark, and M.~Uflacker, ``{Teaching Agile the Agile Way —
  Employing Self-Organizing Teams in a University Software Engineering
  Course},'' in {\em American Society for Engineering Education (ASEE)
  International Forum}, ASEE, 2016.

\bibitem{Salman2015}
I.~Salman, A.~T. Misirli, and N.~Juristo, ``{Are Students Representatives of
  Professionals in Software Engineering Experiments?},'' in {\em 2015 IEEE/ACM
  37th IEEE International Conference on Software Engineering}, vol.~1,
  pp.~666--676, IEEE, may 2015.
\newblock DOI: \texttt{10.1109/ICSE.2015.82}.

\bibitem{Both2014}
T.~Both, ``{I Like , I Wish , What If},'' tech. rep., Stanford D.School, 2014.

\bibitem{leung2001}
W.-C. Leung, ``{How to design a questionnaire},'' {\em Student BMJ}, no.~9,
  pp.~171--216, 2001.

\end{thebibliography}
